\documentclass[10pt,aps,prl,twocolumn,byrevtex,superscriptaddress,showpacs]{revtex4-1}
\usepackage{microtype}
\usepackage{graphics}
\usepackage{amsmath}
\usepackage{amssymb}

\newcommand{\be}{\begin{equation}}
\newcommand{\ee}{\end{equation}}
\newcommand{\ra}{\rightarrow}
\newcommand{\mD}{\mathcal{D}}
\newcommand{\mL}{\mathcal{L}}

\newcommand{\ave}[1]{\langle #1\rangle}

\begin{document}

\title{Nonequilibrium microcanonical and canonical ensembles and their equivalence}

\author{Rapha\"el Chetrite}
\email{Raphael.Chetrite@unice.fr}
\affiliation{\mbox{Laboratoire J.\ A.\ Dieudonn\'e, UMR CNRS 6621, Universit\'e de Nice Sophia-Antipolis, Nice 06108, France}}

\author{Hugo Touchette}
\email{htouchet@alum.mit.edu}
\affiliation{National Institute for Theoretical Physics (NITheP), Stellenbosch 7600, South Africa}
\affiliation{Institute of Theoretical Physics, University of Stellenbosch, Stellenbosch 7600, South Africa}

\date{\today}

\begin{abstract}
Generalizations of the microcanonical and canonical ensembles for paths of Markov processes have been proposed recently to describe the statistical properties of nonequilibrium systems driven in steady states. Here we propose a theory of these ensembles that unifies and generalizes earlier results, and show how it is fundamentally related to the large deviation properties of nonequilibrium systems. Using this theory, we provide conditions for the equivalence of nonequilibrium ensembles, generalizing those found for equilibrium systems, construct driven physical processes that generate these ensembles, and re-derive in a simple way known and new product rules for their transition rates. A nonequilibrium diffusion model is used to illustrate these results.
\end{abstract}


\maketitle

Equilibrium properties of many-particle systems are determined by having recourse to statistical ensembles, such as the microcanonical ensemble for systems with constant energy or the canonical ensemble for systems at constant temperature~\cite{reif1965}. For nonequilibrium systems modeled with Markov processes, similar ensembles can be constructed by defining probability distributions on the trajectories or \emph{paths} of these systems, which are either conditioned on some constraints, as in the microcanonical ensemble, or have the form of a Gibbs distribution, as in the canonical ensemble. Such \emph{nonequilibrium path ensembles} have been considered, in particular, by Evans \cite{evans2004,evans2005a,evans2010} in the context of sheared fluids, and have come to play an important role in recent studies of the glass transition, e.g., in Lennard-Jones liquids \cite{merolle2005,hedges2009,chandler2010}, and dynamical phase transitions in kinetically-constrained models \cite{garrahan2007,garrahan2009,jack2010,speck2011} and quantum systems \cite{ates2012,garrahan2011,genway2012,hickey2012}.

It is known from \cite{evans2004,evans2005a,evans2010} that microcanonical path ensembles can be generated in the long-time limit by a Markov process whose transition rates are those of the original process, modified by a factor taking into account the microcanonical constraint. This result, extended to canonical path ensembles by Jack and Sollich \cite{jack2010b}, is important physically as it shows that  path ensembles can be generated by an effective or \emph{driven} process. However, as of now it is not clear for which systems this result holds and how a microcanonical driven process is related to a canonical one. These problems relate to the issue of ensemble equivalence, which is fundamental in equilibrium statistical mechanics \cite{touchette2011b}.

Here, we present a systematic approach to path ensembles that addresses these problems and show how they are fundamentally related, via the theory of large deviations \cite{touchette2009}, to dynamical fluctuations of nonequilibrium systems. Our theory contains many results obtained before, but also simplifies and generalizes them in many ways. In particular, we are able to treat an important class of nonequilibrium systems not covered in previous studies, namely, diffusions, and to provide a direct proof of known constraint rules satisfied by driven processes \cite{baule2008,baule2010b,simha2008} in addition to derive new ones. From the connection with large deviations, we also state explicit conditions for the equivalence of microcanonical and canonical paths ensembles, similar to those of equilibrium systems, indicating that these ensembles are not always equivalent. To highlight the case of diffusions, we illustrate our results for a driven periodic Langevin equation,  used for example to model manipulated colloidal particles \cite{seifert2012}.

We consider as a general model of physical systems driven in nonequilibrium steady states an ergodic Markov process $X_t$ evolving over a time $t\in [0,T]$. The specific form of this process depends on the system considered. It can be a jump process if one considers discrete-state processes, such as noisy chemical and biological reactions \cite{seifert2012} or interacting particles evolving on a lattice \cite{derrida2007}. Alternatively, it can be a diffusion, or a set of coupled stochastic differential equations in general, if one considers systems such as colloidal particles immersed in liquids \cite{seifert2012}, stochastic thermostated systems \cite{berendsen1984}, or coupled noisy oscillators. In both cases, the effect of external reservoirs and fields can be modeled in terms of boundary conditions or directly at the level of transition rates, drift and diffusion coefficients.

The statistical properties of $X_t$ are determined by its \emph{Markov generator} $L$, which governs the evolution of functions of $X_t$ or by its adjoint $L^\dag$, which governs the evolution of the probability density function (pdf) $p(x,t)$. Formally, we can also characterize $X_t$, following the path integral approach to Markov processes, by the functional pdf $P[x]$ of the paths $\{x(t)\}_{t=0}^T$, which enters in the calculation of expectations of the form
\be
\ave{A_T}=\int \mD [x]\, P[x]\, A_T[x],
\ee
where $\mD[x]$ denotes the path integral element and $A_T[x]$ is a general \emph{observable} of $X_t$. Physical examples of observables include particle and energy currents, shear induced by external forces, or the work performed to drive a process in a nonequilibrium steady state.

In general, one is interested in studying not just the average value of an observable $A_T$, but also its typical (i.e., most probable) value and fluctuations determined by its pdf $P(A_T=a)$. Following the case of equilibrium systems, it is also of interest to study the properties of a process $X_t$ when some observable acting as a constraint is varied. For example, one can study how a fluid initially at equilibrium goes to a nonequilibrium state under shearing \cite{evans2004,evans2005a,evans2010} or how the properties of a glassy system vary as a function of its dynamical activity \cite{merolle2005,hedges2009,chandler2010}. As another example recently considered in \cite{popkov2010,popkov2011,belitsky2013}, one can study how the stationary distribution of a nonequilibrium interacting particle system changes as its current $J_T$ integrated over a time $T$ is observed to be far from its typical value.

In terms of probabilities, these situations are described by conditioning the path pdf $P[x]$ of $X_t$ on the constraint or fluctuation $A_T=a$ observed:
\be
P^a[x]=P[x|A_T=a]=\frac{P[x,A_T=a]}{P(A_T=a)}.
\ee
This defines, similarly to equilibrium, a \emph{microcanonical path ensemble}, also referred to as a \emph{conditional} or \emph{constrained ensemble} \cite{evans2004,evans2005a,evans2010}. Alternatively, we can follow Gibbs and define a \emph{canonical path ensemble} by
\be
P_k[x]=\frac{e^{TkA_T[x]}P[x]}{W_T(k)},\qquad W_T(k)=\ave{e^{TkA_T}},
\ee
which replaces the constraint $A_T=a$ by an exponential involving the parameter $k$ conjugated to $A_T$ \footnote{We use $k$ rather than $\beta$ to make clear that $k$ is not the inverse temperature in general.}. In other works, $P_k$ is also referred to as the \emph{biased}, \emph{tilted} or \emph{$s$-ensemble} \cite{jack2010b,lecomte2005,lecomte2007}. The analogy with equilibrium ensembles is obvious if one views paths as microstates and $A_T$ as the energy. From this, we see that the normalization constant $W_T(k)$ is the analog of the partition function.

Unlike its equilibrium counterpart, the canonical path ensemble does not always have a physical interpretation in terms of heat baths or driving fields. However, from a theoretical point of view, it is interesting to ask, as Gibbs did, whether $P_k$ is equivalent to $P^a$, i.e., whether the properties of a system (equilibrium or stationary) in the canonical ensemble reproduce those of the microcanonical ensemble. This equivalence is known to hold for many equilibrium systems \cite{touchette2011b} and justifies using the canonical ensemble for actual calculations because of its simpler, unconstrained form. The same problem applies here: if we can demonstrate that $P_k$ and $P^a$ are equivalent, then any calculations of typical quantities involving $P^a$ can be done with $P_k$. Proving equivalence also provides a way to physically obtain the canonical ensemble from the microcanonical ensemble by Gibbs conditioning.

Below, we show that equivalence holds in the stationary limit where $T\ra\infty$, under some conditions related to the fluctuations of $A_T$. Moreover, we show that there exists a Markov process with generator $L_k$, different from $L$, whose path pdf is equivalent to $P^a$ and $P_k$ in the same limit. Physically, this means that \emph{the microcanonical and canonical path ensembles are generated by a specific process, called as before the driven process, involving additional forces compared to those driving $X_t$}.

This is essentially the result of Evans mentioned earlier, with the difference that we now treat the problem of equivalence explicitly. Moreover, we obtain this result for general Markov processes, as well as a general class of observables that are i) integrated in time and ii) depend on the state of the process and its transitions over time. For jump processes, these observables take the form
\be
A_T=\frac{1}{T}\int_0^T f(X_t)dt+\frac{1}{T}\sum_{0\leq t\leq T: \Delta X_t\neq 0} g(X_{t^-},X_{t^+}),
\label{eqobs1}
\ee
where $f$ and $g$ are arbitrary functions and the sum is over all jumps $\Delta X_t=X_{t^+}-X_{t^-}\neq 0$ of the process.
For diffusion processes, observables satisfying i) and ii) above are of the form
\be
A_T=\frac{1}{T}\int_0^T f(X_t)\, dt+\frac{1}{T}\int_0^T g(X_t)\circ dX_t,
\label{eqobs2}
\ee
where $\circ$ denotes the Stratonovich integral \cite{stroock1979}. These observables are very general: they include as special cases all the quantities mentioned so far (currents, shear, activity) in addition to quantities such as the integrated work, heat, and entropy production considered in the context of stochastic thermodynamics \cite{sekimoto2010}.

The fundamental assumption underlying our results is that \emph{the observable $A_T$ satisfies a large deviation principle} \cite{touchette2009}. This means that the stationary pdf of $A_T$ scales in the limit $T\ra\infty$ as
\be
P(A_T=a)= e^{-TI(a)+o(T)},
\ee
where $I(a)$ is a function, called the \emph{rate function}, that does not depend on $T$. This function plays a central role in equilibrium and nonequilibrium statistical physics \cite{touchette2009}, as it characterizes the typical values and fluctuations of $A_T$. In large deviation theory, it is obtained from the dominant eigenvalue $\Lambda_k$ of the so-called \emph{tilted operator} $\mL_{k}$, whose form depends on the observable considered. For observables having the jump form of (\ref{eqobs1}), we have 
\be
\mL_{k}=Le^{kg} +kf,
\label{eqtiljp}
\ee
whereas for diffusion observables defined in (\ref{eqobs2}), we have \footnote{This follows by combining the Feynman-Kac and Girsanov formulae \cite{stroock1979}.}
\be
\mL_{k}=\hat F(\nabla+kg)+(\nabla+kg)\frac{D}{2}(\nabla+kg)+kf,
\label{eqdiftg}
\ee
where $\hat F$ is the corrected drift entering in the symmetrized form of the generator of the stochastic differential equation considered and $D$ is its diffusion matrix \footnote{All terms in these expressions operators: $L e^{kg}$ is the operator with components $(L e^{kg})(x,y)=L(x,y) e^{kg(x,y)}$, while $kf$ is a diagonal operator with components $k f(x) \delta(x-y)$.}.

Assuming a large deviation principle for $A_T$, we can already solve the equivalence problem by appealing to general results of \cite{touchette2011b} stated for equilibrium ensembles, but which equally apply to nonequilibrium ensembles. The definition of equivalence used here is the following: we say that two path pdfs $P$ and $Q$ are \emph{asymptotically equivalent} if
\be
\lim_{T\ra\infty} \frac{1}{T}\ln \frac{P[x]}{Q[x]}=0
\label{eqequiv1}
\ee
almost everywhere with respect to $P$. Essentially, this means that $P$ and $Q$ are equivalent if they are equal up to subexponential terms in $T$ and for almost all paths.

This notion of equivalence is slightly stronger than the one used in \cite{touchette2011b}, but also implies that the typical values of any observable obtained under $P$ as $T\ra\infty$ are the same as those obtained under $Q$ in this limit. Moreover, the equivalence of $P^a$ and $P_k$ in the sense of (\ref{eqequiv1}) turns out to be determined, as in \cite{touchette2011b}, by the convexity of the rate function $I(a)$ as follows:

(i) If $I(a)$ is convex at the point $a$, then there exists a $k$ such that $P_k$ is asymptotically equivalent to $P^a$, which means physically that the two ensembles describe the same long-time stochastic dynamics. Moreover, if $I(a)$ is differentiable, then equivalence holds for $k=I'(a)$, providing a generalization of the temperature-entropy relation.

(ii) If $I(a)$ is nonconvex at $a$, then there is no $k$ for which $P_k$ is asymptotically equivalent with $P^a$. In this case, the microcanonical process obtained by conditioning on $A_T=a$ has no canonical counterpart for any $k$.

The proof of these results, as well as the discussion of a third, more technical case of equivalence, known as partial equivalence, are deferred to a longer paper \footnote{R. Chetrite, H. Touchette, in preparation.}. What is important to note here is that the equivalence of path ensembles is not guaranteed -- it depends on the convexity of the rate function $I(a)$, in a way similar to equilibrium ensembles, whose equivalence depends on the concavity of the thermodynamic entropy \cite{touchette2011b}. In most works on path ensembles, this equivalence is assumed.

With this result, we now turn to the problem of generating or representing the microcanonical and canonical path ensembles by a Markov process. A priori, it is not clear that such a Markov representation exists, since neither $P^a$ nor $P_k$ is Markovian: the former has a conditioning that depends on the future time $T$, while the latter involves the constant $W_T(k)$ which breaks the multiplicative structure of the path pdf. However, as mentioned before, it is possible to construct a Markov process with generator $L_k$, whose path pdf is \emph{asymptotically} equivalent to $P_k$, in the same sense as above, and which, in the case of ensemble equivalence, must therefore also be equivalent to $P^a$.

To construct this Markov process, we consider the following \emph{similarity transform} of $L$:
\be
L^h=h^{-1}Lh-h^{-1}(Lh),
\ee
which is a generalization of Doob's $h$-transform \cite{doob1984,chung2005,chetrite2011} defined for positive functions $h$ \footnote{Multiplications in this expression have the same sense as in $\mL_k$, except for $(Lh)$ which stands for the application or action of $L$ on $h$.}. 
For the purpose of defining $L_k$, the main property of $L^h$ that we need to know is that it is a Markov generator, and so defines a Markov process which has an explicit probability ratio with respect to the original path pdf $P[x]$. Combining the expression of this ratio, found in \cite{chetrite2011}, with $\mL_k$, it is then possible to construct the equivalent Markov process with generator $L_k$ and path pdf $Q[x]$ in such a way that the limit of (\ref{eqequiv1}) with $P/Q$ replaced by $Q/P_k$ vanishes. The full derivation is deferred again to \cite{Note4}. 

The final result is that $L_k$ is unique and is obtained by applying the $h$-transform to $\mL_k$ with $h=r_k$, where $r_k$ is the (right) eigenvector of $\mL_k$ associated with the dominant eigenvalue $\Lambda_k$, so that
\be
L_k=\mL_{k}^{r_k}.
\ee
Noticing that the similarity does not act on the diagonal part of $\mL_{k}$ involving $f$, we can also write
\be
L_k=r_k^{-1} \left.\mL_{k}\right|_{f=0}\, r_k +kf-\Lambda_k.
\label{eqmainres}
\ee
This explicit expression for the generator of the Markov process asymptotically equivalent to the canonical path ensemble is the main result of this paper. It applies to any Markov processes, including jump processes and diffusions, as well as any observables of these processes, provided that they satisfy a large deviation principle.

Before discussing a specific application, it is instructive to show that (\ref{eqmainres}) recovers previous results obtained for physical jump processes characterized by jump rates $W(x,y)$ from state $x$ to $y$. In this case, we obtain from (\ref{eqmainres}) and (\ref{eqtiljp}) that the rates $W_k(x,y)$ of the driven process associated with the canonical path ensemble have the form
\be
W_k(x,y)=r_k^{-1}(x)W(x,y)e^{kg(x,y)}r_k(y),
\ee
which is the result obtained by Jack and Sollich \cite{jack2010b} and essentially also Evans \cite{evans2004,evans2005a,evans2010}. From this, we immediately obtain
\be
W_k(x,y)W_k(y,x)=W(x,y)W(y,x)e^{k[g(x,y)+g(y,x)]},
\ee
which reduces to
\be
W_k(x,y)W_k(y,x)=W(x,y)W(y,x)
\ee
if $g$ is an antisymmetric function. This identity relating the driven rates to the original rates of $X_t$ is one of the \emph{product constraint rules} recently derived in \cite{baule2008,baule2010b,simha2008} by more involved methods. Another constraint found by these authors relates the escape rate $\nu(x)=(W1)(x)$ of $X_t$ to the escape rate $\nu_k(x)=(W_k1)(x)$ of the driven process:
\be
\nu(x)-\nu(y)=\nu_k(x)-\nu_k(y).
\ee
In our formalism, this \emph{exit rate constraint} is simply obtained from (\ref{eqmainres}) for $f=0$ and any function $g$. In the case where $f\neq 0$, we obtain the more general identity,
\be
\nu_k(x)-\nu(x)+kf(x)=\Lambda_k.
\ee

To see how these results can be applied beyond jump processes, consider now a general $d$-dimensional diffusion having the (It\^o) form
\be
dX_t=F(X_t)dt+\sigma(X_t) dW_t,
\ee
where $F$ is a $d$-dimensional force field, $\sigma$ is a $d\times d$ noise matrix, and $dW_t$ is a $d$-dimensional Wiener noise \cite{stroock1979}. Assuming that this process is ergodic, we can derive from its generator and (\ref{eqmainres}) that the driven process generating the canonical path ensemble of $X_t$ is a diffusion with the same noise as above but with the modified force
\be
F_k=F+D(kg+\nabla \ln r_k),
\label{eqmodf}
\ee
where $D=\sigma \sigma^T$ is the diffusion matrix. If the canonical path ensemble is equivalent to the microcanonical path ensemble, then the driven diffusion with the force $F_k$ also generates the process $X_t$ conditioned on the fluctuation $A_T=a$. As a consequence of this result, we see that the driven process associated with an equilibrium system \footnote{An equilibrium system has a stationary distribution $\rho$, drift, and diffusion related by $F=\frac{D}{2}\nabla \ln \rho+\frac{1}{2}\nabla D$.} is itself an equilibrium system if $g$ is gradient and is otherwise modified to a nonequilibrium system not satisfying detailed balance. If $g$ is gradient, we also obtain from (\ref{eqmodf}) the identity
\be
\nabla\times (D^{-1} F_k)=\nabla\times(D^{-1}F)
\label{eqrot1}
\ee
for 2- and 3-dimensional systems \footnote{A similar formula involving exterior derivatives holds for dimensions greater than 3.}, which can be seen as a diffusive version of the jump process constraint rules.

To illustrate these results, let us consider the forced periodic diffusion evolving according to
\be
\dot \theta(t)=\gamma -V'(\theta(t))+\xi(t),\quad \theta\in[0,2\pi)
\label{eqmodel1}
\ee
where  $V=V_0\cos(\theta)$, $\gamma$ is a constant frequency driving the system in a nonequilibrium steady state breaking detailed balance, and $\xi(t)$ is Gaussian a white noise. This model arises in many physical applications, e.g., ion conduction, noisy Josephson junctions \cite{risken1996}, in addition to Brownian motors and colloidal particles \cite{seifert2012}, and serves as an important model for testing new ideas and results about nonequilibrium systems; see, e.g., \cite{blickle2007,gomez2009,nemoto2011b,*nemoto2011}. 

\begin{figure}[t]
\resizebox{3.4in}{!}{\includegraphics{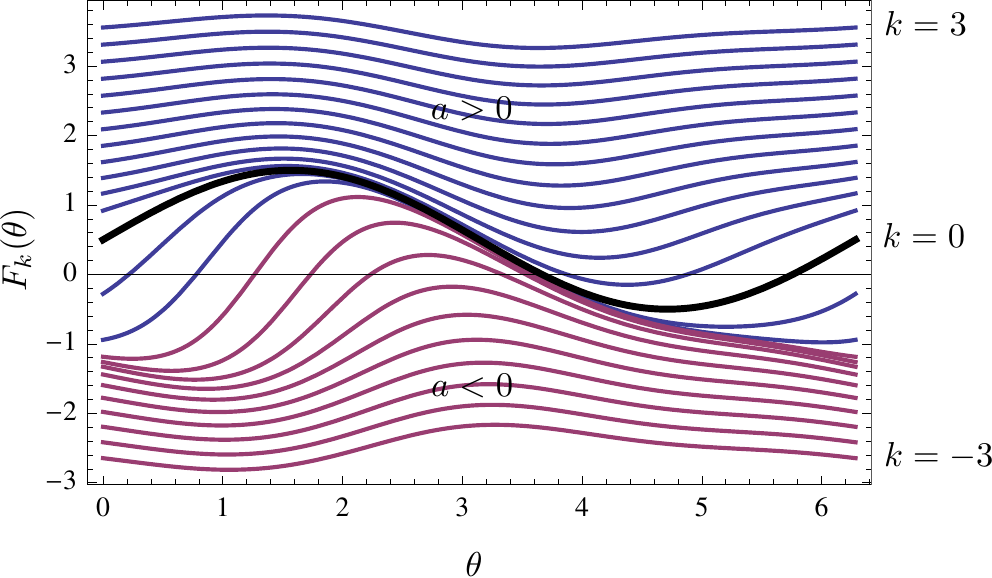}}
\caption{(Color online) Force $F_k(\theta)$ of the driven Markov process associated with the canonical version of the periodic diffusion with $V_0=1$ and $\gamma=0.5$. The different curves are obtained for $k\in [-3,3]$ in steps of $0.25$. The force of the base process corresponding to $k=0$ is shown in black. Blue curves yield positive stationary currents, $A_T=a>0$, while purple curves yield negative stationary currents.}
\label{fig1}
\end{figure}

An important observable of this system is the velocity averaged over a time $T$:
\be
A_T=\frac{1}{T}\int_0^T d\theta(t)=\frac{\theta(T)-\theta(0)}{T},
\ee
which is proportional to the mean current. This is an observable having the form (\ref{eqobs2}) with $f=0$ and $g=1$, for which the tilted generator is, according to (\ref{eqdiftg}),
\be
\mL_k=(\gamma-V'+k)\frac{d}{d\theta}+\frac{1}{2}\frac{d^2}{d\theta^2}+\frac{k^2}{2}+k(\gamma-V').
\ee
The dominant eigenfunction $r_k(\theta)$ of this non-hermitian operator is easily found by expanding the eigenvalue equation in Fourier modes. The effective force $F_k(\theta)$ of the driven process obtained from $r_k(\theta)$ is shown in Fig.~\ref{fig1} for $V_0=1$ and $\gamma=0.5$, in which case the stationary (long-time) value of the current $A_T$ is $a^*\approx0.176$.

We see from the plot that the effective force of the driven process generating the canonical path ensemble is always periodic but changes in a non-trivial way with $k$. For $k>0$, $F_k$ drives the system to a positive stationary value of the current greater than $a^*$, while for $k<0$, $F_k$ drives the system to positive currents for $k$ down to some value $k=0.5$, below which the current becomes negative in the stationary limit. Thus, by choosing $k$ properly, we can force the system to assume any stationary, nonequilibrium current. In the limit $k\ra\pm\infty$, the constant $k^2$ term in the tilted generator dominates and leads to $F_k\approx k$, which means physically that the effective dynamics behind large currents in the canonical ensemble is a simple potential-free diffusion with a constant drive in the direction of the current. Since the rate function of $A_T$ is known to be convex for this model, this also means that the periodic diffusion (\ref{eqmodel1}) is equivalent to a driven diffusion with constant frequency when it is conditioned microcanonically on large currents.

To conclude, we have presented a general theory of microcanonical and canonical path ensembles describing the typical states and fluctuations of Markovian systems driven in nonequilibrium steady states. This theory enables one to determine, similarly to equilibrium systems, when these two ensembles are equivalent (that is, when they describe the same stationary states) and shows how the nonequilibrium dynamics associated with each ensemble is generated physically by an effective Markov process representing a driven nonequilibrium system. It also shows in a clear and rigorous way that the properties of this driven process are intimately related to the large deviation properties of nonequilibrium systems. In the future, we will provide proofs of the results presented here, in addition to investigate new problems arising from these results---in particular, the significance of the gauge-like rotational constraint (\ref{eqrot1}), the derivation of driven processes for systems having nonconvex rate functions, for which equivalence of ensembles does not hold, and the use of driven processes in large deviation simulations, an important tool in studies of nonequilibrium systems.

\begin{acknowledgments}
H.T. is grateful to Stefano Ruffo and Thierry Dauxois for supporting a visit to ENS Lyon with the ANR grant LORIS (ANR-10-CEXC-010-01).
\end{acknowledgments}

\bibliography{masterbibmin}

\end{document}